\documentclass[reprint,aip]{revtex4-1}
\usepackage[english]{babel}
\usepackage{geometry,amsmath,graphicx,hyperref}
\newcommand{\tmem}[1]{{\em #1\/}}
\newcommand{\tmop}[1]{\ensuremath{\operatorname{#1}}}
\newcommand{\tmtextbf}[1]{\text{{\bfseries{#1}}}}
\newcommand{\tmtextit}[1]{\text{{\itshape{#1}}}}
\newcommand{\tmtexttt}[1]{\text{{\ttfamily{#1}}}}
\newcommand{\nonconverted}[1]{\mbox{}}

\begin{document}

\title{Simulation of neutral beam current drive on EAST tokamak}

\author{Youjun Hu}
\email{yjhu@ipp.cas.cn}
\affiliation{Institute of Plasma Physics, Chinese Academy of Sciences, Hefei 230031, China}
\author{Xingyuan Xu}
\affiliation{Institute of Plasma Physics, Chinese Academy of Sciences, Hefei 230031, China}
\author{Yunchan Hu}
\affiliation{Institute of Plasma Physics, Chinese Academy of Sciences, Hefei 230031, China}
\author{Kaiyang He}
\affiliation{Institute of Plasma Physics, Chinese Academy of Sciences, Hefei 230031, China}
\author{Jinfang Wang}
\affiliation{Institute of Plasma Physics, Chinese Academy of Sciences, Hefei 230031, China}

\begin{abstract}
  Neutral beam current drive (NBCD) on the EAST tokamak is studied by using
  Monte-Carlo test particle code \tmtexttt{TGCO}. Phase-space structure of the
  steady-state fast ion distribution is examined and visualized. We find that
  trapped ions carry co-current current near the edge and counter-current
  current near the core. However, the magnitude of the trapped ion current is
  one order smaller than that of the passing ions. Therefore their
  contribution to the fast ion current is negligible ($1\%$ of the fast ion
  current). We examine the dependence of the fast ion current on two basic
  plasma parameters: the plasma current $I_p$ and plasma density $n_e$. The
  results indicate that the dependence of fast ion current on $I_p$ is not
  monotonic: with $I_p$ increasing, the fast ion current first increases and
  then decreases. This dependence can be explained by the change of trapped
  fraction and drift-orbit width with $I_p$. The fast ion current decreases
  with the increase of plasma density $n_e$. This dependence is related to the
  variation of the slowing-down time with $n_e$, which is already well known
  and is confirmed in our specific situation. The electron shielding effect to
  the fast ion current is taken into account by using a fitting formula
  applicable to general tokamak equilibria and arbitrary collisionality
  regime. The dependence of the net current on the plasma current and density
  follows the same trend as that of the fast ion current.
\end{abstract}

{\maketitle}

\

\section{Introduction}

Neutral beam injection (NBI) is widely used in tokamaks and stellarators for
heating
plasma{\cite{Scoville2019,Hu2015,Schneider_2011,ASUNTA201533,Heidbrink_2009,Buttery2019}}.
Besides heating, the fast ions resulting from NBI can also drive electric
current in plasmas{\cite{park2009,Oikawa2008,kraus2011,Mulas_2023}}. To model
this current, one needs to calculate the steady-state distribution of fast
ions, and then integrate it to get the current. One method of doing this
calculation is to use the Monte-Carlo method to sample NBI fast ion source and
following the guiding-center/full trajectories of fast ions, taking into
account of their collisions with the background plasmas. Fusion community has
developed many computer codes doing this kind of simulations, among which are
\tmtexttt{NUBEAM}{\cite{pankin2004}}, \ \tmtexttt{OFMC}{\cite{Tani1981}},
\tmtexttt{ASCOT}{\cite{HIRVIJOKI2014}}, \tmtexttt{ORBIT}{\cite{White_2010}},
\tmtexttt{SPIRAL}{\cite{Kramer2013}}, and many
others{\cite{PFEFFERLE20143127,yxu2019,he2020b,Wang_2021}}. An advantage of
this method is that it can readily take into account the real space effects,
such as finite orbit width and edge loss, which are usually approximately
treated in analytical models{\cite{Taguchi_1996,hirshman:1238}} and some
velocity grid based Fokker-Planck codes.

The fast ion steady-state distribution is of interest not only to current
drive problem but also to many research topics such as the interactions
between fast ions and MHD modes and turbulence. Some mechanisms in the
interaction may depend on delicate phase-space structure of fast ions.
Therefore an accurate fast ion distribution function is of practical
importance and the phase space structure needs to be more thoroughly studied
and visualized than that has been done previously. In this paper, we carefully
examine the phase-space structure of the steady-state fast ion distribution.
We find that trapped ions carry co-current (relative to the main plasma
current direction) current near the edge and counter-current current near the
core. The magnitude of the trapped ion current is one order smaller than that
of the passing ions. Therefore their contribution to the fast ion current is
very small ($1\%$ of the fast ion current).

We consider co-current NBI (all the four beams on EAST
tokamak{\cite{Wan_2019}} are now in co-current direction). In order to operate
EAST for longer pulse, there are interests in getting better neutral beam
current drive (NBCD) efficiency by operating in optimized parameter regimes.
In this work, based on realistic EAST configuration and plasma profiles, we
examine the dependence of NBCD on two basic plasma parameters, namely the
plasma current $I_p$ and plasma density $n_e$, using a Monte-Carlo test
particle code \tmtexttt{TGCO}{\cite{youjunhu2021,Xu_2020}}. The results
indicate that, with the plasma current increasing, the fast ion current first
increases and then decreases. With the plasma density increase, the fast ion
current decreases. The $n_e$ dependence is related to the variation of the
slowing-down time with $n_e$, which is already well known and is confirmed in
our specific situation. The $I_p$ dependence is not well known and need some
explanations. We note that the drift orbit width of a fast ion is inversely
proportional to $I_p${\cite{wesson2004}}. This means that the fast ion
confinement improves with the increase of $I_p$. This may imply that NBCD
efficiency also improve with the increase of $I_p$. However, the simulations
in this paper indicate this is not the complete picture: the NBCD efficiency
turns out to decrease with the increase of the plasma current in the larger
$I_p$ regime. This trend is found to be due to the fast ion trapped fraction
increasing with the increase of $I_p$. As mentioned above, the trapped
particle carries nearly zero current. Larger fraction of trapped particles
implies smaller fast ion current.

To get the net current, one needs to take into account the electron shielding
effect, i.e., the current carried by electrons due to their response to the
presence of fast ions. This generally requires to solve the steady-state
Fokker-Planck equation for electrons with additional collision term
corresponding to the electron collision with the fast ions. For current drive
problem, we are interested in the first (parallel) moment of the electron
distribution function. Then, making use of the self-adjoint property of the
linearized collision operator, the electron response to arbitrary fast ion
sources can be obtained by using the Green function method
{\cite{lin-liu1997,sauter1999,Honda_2012,youjunhu2012}}. The final results of
these studies are usually some fitting formulas for the ratio of net current
to the pure fast ion current{\cite{lin-liu1997,Honda_2012}}. The present work
uses these fitting formulas to include the electron shielding effect. The
electron shielding model used in this work is a general model applicable to
arbitrary collisionality regime and general tokamak flux surface
shapes{\cite{youjunhu2012,sauter1999,Honda_2012,lin-liu1997}}. Our results
show that the $I_p$ and $n_e$ dependence of the net current is similar to that
of the fast ion current.

\section{Simulations}

\subsection{Plasma equilibrium configuration and profiles}

\begin{figure}[h]
  \resizebox{0.95\columnwidth}{!}{\includegraphics{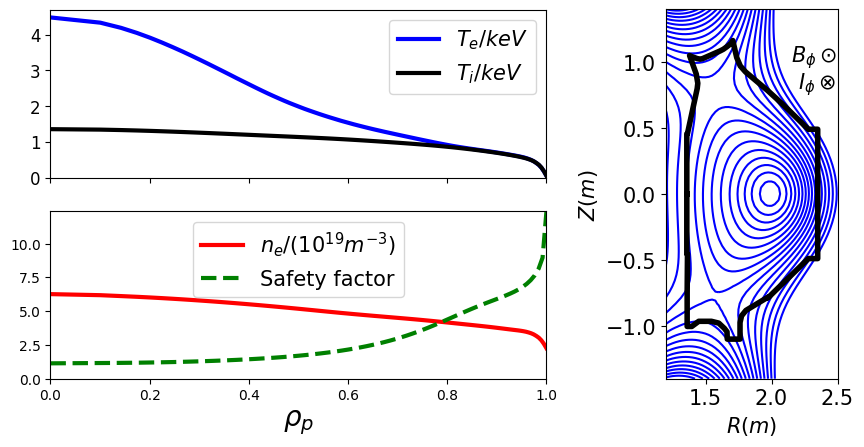}}
  \caption{\label{17-2-22-p1m}Left panel: profiles of electron number density
  $n_e$ , electron temperature $T_e$, ion temperature $T_i$, and safety factor
  $q$. Right panel: magnetic configuration. Directions of the current and
  toroidal magnetic field are indicated in the figure. This is a
  high-$\beta_p$ discharge with low plasma current $I_p = 360 \tmop{kA}$,
  $q_{95} = 7.6$, $q_{\tmop{axis}} = 1.16$, and $B_{\phi \tmop{axis}} = - 2.2
  T$. (All visualizations in this paper were created by using
  Numpy{\cite{harris2020array}}, Scipy{\cite{2020SciPy-NMeth}}, and
  Matplotlib{\cite{Hunter:2007}}, which are open source Python libraries.)}
\end{figure}

The simulations are performed for the EAST tokamak, which is a superconducting
tokamak with a major radius $R_0 = 1.85 m$, minor radius $a \approx 0.45 m$,
typical on-axis magnetic field strength $B_{\tmop{axis}} \approx 2.2 T$ and
plasma current $I_p \approx 0.5 \tmop{MA}${\cite{Wan_2017,Wan_2019}}. Figure
\ref{17-2-22-p1m} plots the magnetic configuration and plasma profiles used in
this work, which were reconstructed by the EFIT code{\cite{lao1985}} from EAST
tokamak discharge \#101473@4.5s with constrains from experiment
diagnostics{\cite{huyunchan2023}}. The toroidal plasma current is in $+
\hat{\mathbf{\phi}}$ of cylindrical coordinate $(R, \phi, Z)$. The radial
coordinate $\rho_p $used in this paper is the square root of the normalized
poloidal magnetic flux: $\rho_p = \sqrt{(\Psi - \Psi_0) / (\Psi_b - \Psi_0)}$,
where $\Psi = R A_{\phi}$ is the poloidal flux function, $A_{\phi}$ is the
toroidal component of the magnetic vector potential, $\Psi_0$ and $\Psi_b$ are
the values of $\Psi$ at the magnetic axis and last-closed-flux-surface,
respectively.

In this work, we assume a Deuterium plasma with Carbon impurities, with the
effective charge number of background ions being $Z_{\tmop{eff}} = 2.23$
across the entire plasma. Simulations in this work are performed by using
\tmtexttt{TGCO} code{\cite{youjunhu2021,Xu_2020}}, which models neutral beam
ionization, slowing-down, collision transport, and edge loss of the resulting
fast ions. We consider Deuterium NBI with full energy $E_{\tmop{full}} = 65
\tmop{keV}$, and the particle number ratio between full, half, and 1/3 energy
being $80\%: 14\%: 6\%$ after the beam leaving from the neutralization vessel.
The beam power after leaving the neutralization vessel is fixed at 1MW. We
consider a reference case where NBI tangential radius is $R_{\tan} = 1.26 m$
and injects in the co-current direction. As a comparison, we also consider a
modified scenario where the tangential radius is changed to $R_{\tan} = 0.731
m$.

\subsection{Fast ion birth distribution}

The neutral beam ionization is modeled by the Monte-Carlo
method{\cite{pankin2004,youjunhu2021}}. Typical number of Monte-Carlo markers
initially loaded in the simulations is $1 \times 10^5$. The beam stopping
cross section data used in the simulation are from Ref. {\cite{Suzuki_1998}},
which includes the charge exchange with thermal and impurity ions, impact
ionization by electrons, thermal and impurity ions, and the multi-step
ionization involving excitation states of neutrals. Beam ionization outside of
the last-closed-flux-surface (LCFS) is ignored in the simulations.

\begin{figure}[h]
  \resizebox{0.95\columnwidth}{!}{\includegraphics{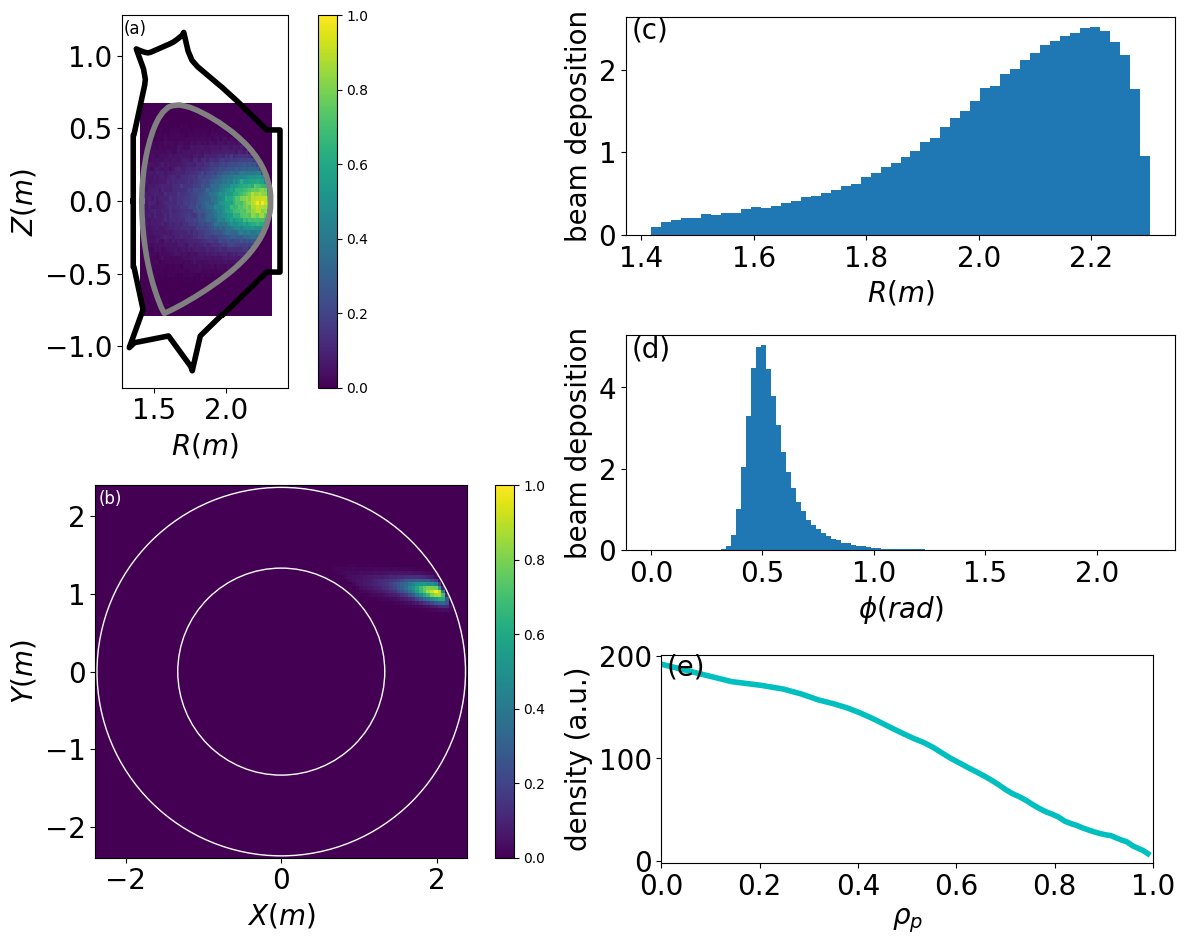}}
  
  \
  \caption{\label{22-10-31-p3}Neutral beam ionization density in the poloidal
  plane (a) and toroidal plane (b). One-dimensional histogram along $R$ and
  $\phi$ are also shown in (c) and (d). Panel (e) shows the density profile
  along the minor radius (averaged over the poloidal and toroidal direction).
  This is for the neutral beam injection with $R_{\tan} = 1.26 m$ and
  $E_{\tmop{full}} = 65 \tmop{keV}$ in EAST discharge \#101473@4.5s. The while
  lines in (b) correspond to the first (inner and outer) wall. The magnetic
  axis is at $R = 1.9 m$. }
\end{figure}

Figure \ref{22-10-31-p3}(a) and (b) plot the fast ion density in the poloidal
plane (averaged over the toroidal direction) and in the toroidal plane
(averaged over the vertical direction), respectively. Figure
\ref{22-10-31-p3}(c) and (d) plot the 1D histogram of the fast ions along $R$
and $\phi$, respectively. The results indicate most fast ions are born near
the low-field-side. The particle and power shine-through fraction in this case
is 3.8\% and 4.1\%, respectively. Figure \ref{22-10-31-p3}(e) plots the
deposition density profile along the minor radius, which shows that the
density reach its maximum at the magnetic axis, i.e., the beam deposition is
on-axis.

\begin{figure}[h]
  \resizebox{0.9\columnwidth}{!}{\includegraphics{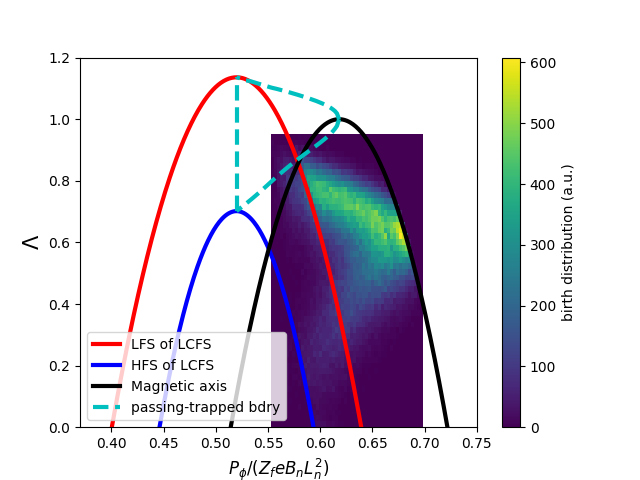}}
  \caption{\label{22-10-18-2}NBI fast ion birth distribution shown in the
  $(P_{\phi}, \Lambda)$ plane. Also shown are the passing trapped boundary,
  the magnetic axis, high-field-side, and the low-field side of
  last-closed-flux-surface (LCFS) for fast ions of $E = 65 \tmop{keV}$. Note
  that the pass-trapped boundary does not depend on the kinetic energy. Here
  $B_n = 1 T$ and $L_n = 1 m$.}
\end{figure}

\begin{figure}[h]
  \resizebox{0.8\columnwidth}{!}{\includegraphics{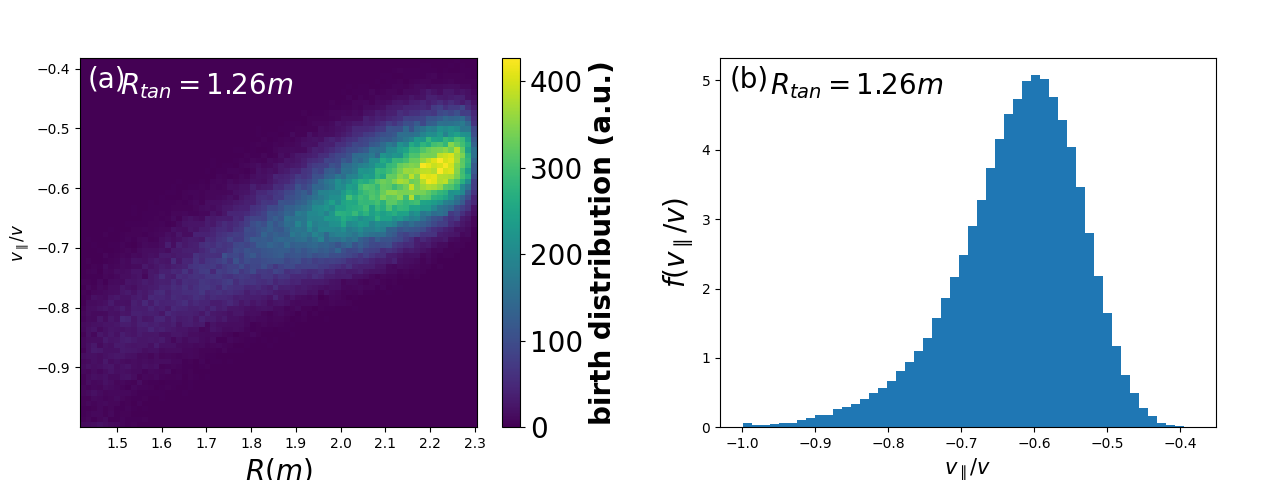}}
  \caption{\label{23-7-15-e1}NBI fast ion birth distribution histogram in $(R,
  v_{\parallel} / v)$ (a) and in $v_{\parallel} / v$ (b).}
\end{figure}

It is often useful to use $(P_{\phi}, \Lambda)$ coordinates to describe the
phase space, where $\Lambda = \mu B_{\tmop{axis}} / E$, $\mu$ is the magnetic
moment, $E$ is the kinetic energy, and $P_{\phi}$ is the canonical toroidal
angular momentum. In these coordinates, it is easier to classify orbit types.
For example, whether a particles is passing or trapped can be readily
identified. Figure \ref{22-10-18-2} plots the fast ion birth distribution in
$(P_{\phi}, \Lambda)$ plane, which indicates that most of the fast ions are
passing particles (the region within the dashed line triangle is the trapped
region). The fraction of trapped fast ions is $0.6\%$ for this case. Figure
\ref{23-7-15-e1} plots the birth distribution over the pitch $v_{\parallel} /
v$, which shows that the dominant pitch is around $- 0.6$.

\subsection{Fast ion steady-state distribution}

Fast ions born from the beam ionization are transformed to guiding-center
space and the guiding-center drifts are followed by using the 4th order
Runge-Kutta scheme in the cylindrical coordinates $(R, \phi, Z)$. A fast ion
is considered lost/thermalized when it touches the first wall or when it slows
down to the energy of $2 T_i (0)$, where $T_i (0)$ is the thermal ion
temperature at the magnetic axis. Charge exchange loss of fast ions
{\cite{Kramer_2020}} is not included in the simulations. The loop voltage is
well controlled to be near zero during the flat-top phase, indicating fully
non-inductive. Therefore, the toroidal electric field is not included in the
simulation. The collision model used in this work includes the effect of
slowing-down, pitch-angle scattering, and energy diffusion (the details are
provided in Appendix \ref{22-11-2-a1}).

The finite Larmor radius (FLR) effect is included when (1) checking whether a
fast ion touches the wall and (2) depositing guiding-center markers to spatial
gridpoints to compute the distribution moment. The FLR effect is not included
when pushing guiding-center trajectories since it has negligible effects.

In order to have a steady state on the slowing-down time scale, we need to
include a continuous beam source. The method of including the continuous beam
source is given in Appendix \ref{22-11-1}.

\begin{figure}[h]
  \resizebox{0.9\columnwidth}{!}{\includegraphics{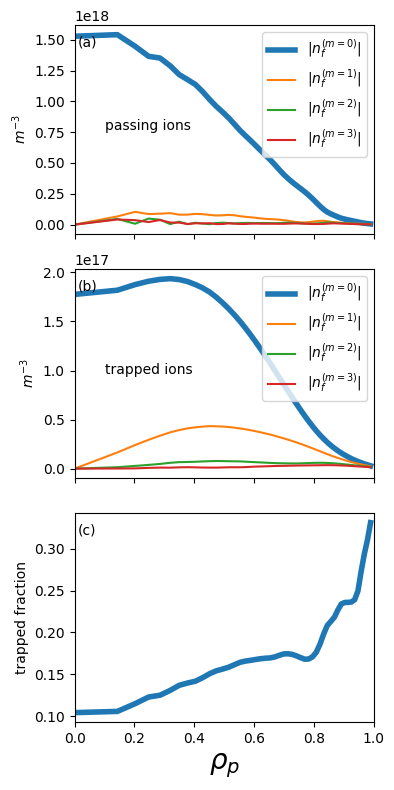}}
  \caption{\label{23-5-10-p3}Radial profiles of various poloidal Fourier
  harmonics of passing (a) and trapped (b) fast ion densities. Shown in (c) is
  the radial profile of trapped fast ion fraction. The straight-field-line
  poloidal angle is used when doing the poloidal Fourier expansion.}
\end{figure}

Figure \ref{23-4-23-p1} plots the steady-state fast ion density distribution
in the poloidal plane (left-panel) and in the toroidal plane (right-panel).
The results indicate that the fast ions density is roughly poloidally uniform
and toroidally uniform for the 1MW beam power. Note that the fast ion source
is neither poloidally uniform nor toroidally uniform. The steady state fast
ion distribution is not guaranteed to be poloidally uniform or toroidally
uniform. It depends on the magnitude of beam power. For the 1MW beam power
considered here, the non-uniformity in the poloidal direction and toroidal
direction is negligible. The poloidal uniformity is further confirmed in
Figure \ref{23-5-10-p3}, which plots the radial profiles of various poloidal
harmonics of the fast ion density. The results show that the $m = 0$ harmonic
is dominant ($m$ is the poloidal mode number).

\begin{figure}[h]
  \resizebox{0.95\columnwidth}{!}{\includegraphics{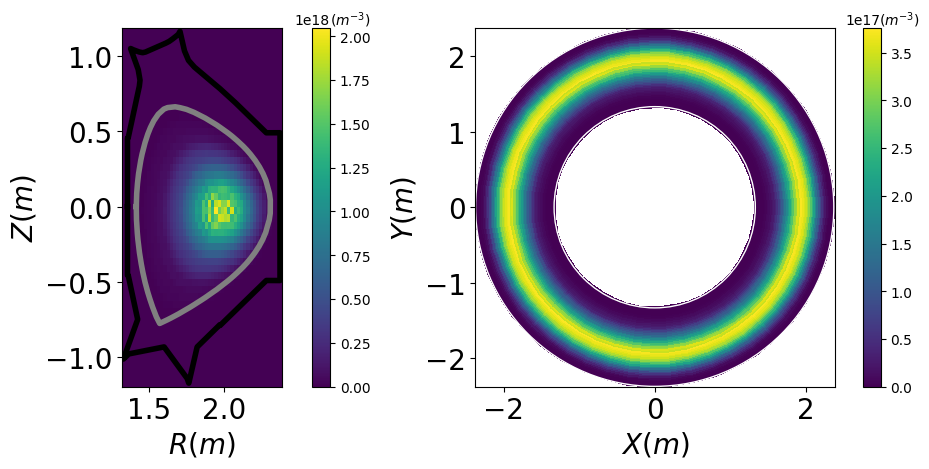}}
  \caption{\label{23-4-23-p1}Fast ion density distribution in the poloidal
  plane (averaged over the toroidal direction) and toroidal plane (averaged
  over the vertical direction).}
\end{figure}

In Fig. \ref{23-5-10-p3}, we distinguish between trapped and passing
particles. The radial profile of the trapped particle fraction is also shown
is Fig. \ref{23-5-10-p3}c. The total trapped particle fraction in the
steady-state distribution is $17\%$, which is significantly different from
that in the birth distribution $(0.6\%)$. The pitch angle scattering can
scatter particles between passing and trapped region of the phase space.
Hence, it is expected the trapped fraction may be changed by collisions.

\begin{figure}[h]
  \resizebox{0.9\columnwidth}{!}{\includegraphics{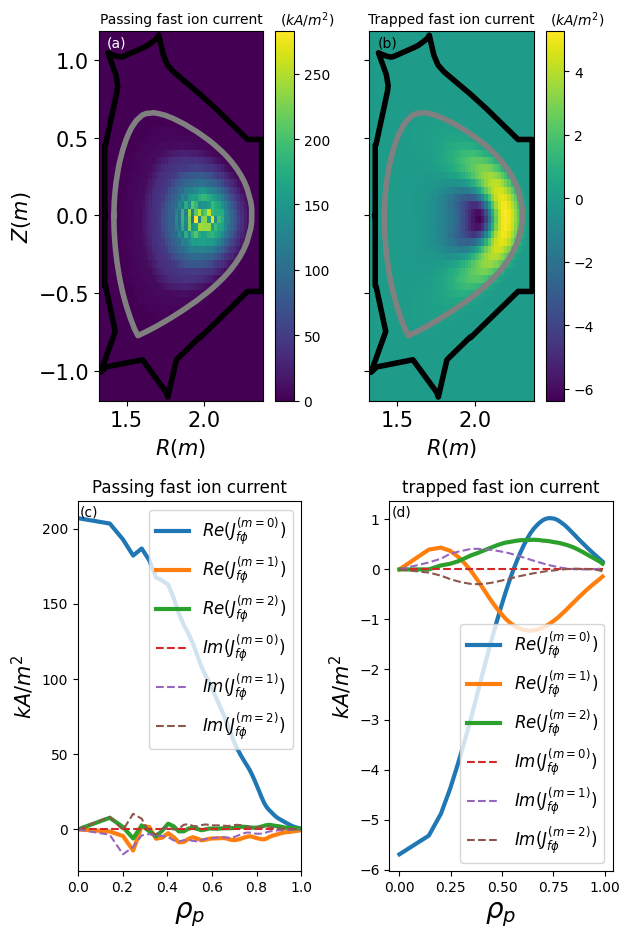}}
  \caption{\label{23-5-10-p4}Distribution of passing (a) and trapped (b) fast
  ion toroidal current densities in the poloidal plane. Shown in (c) and (d)
  are the radial profiles of various poloidal Fourier harmonics of passing (c)
  and trapped (d) fast ion current densities.}
\end{figure}

Figures \ref{23-5-10-p4}a-b plot the fast ion toroidal current density
distribution in the poloidal plane. Here we distinguish the current carried by
passing particles and that carried by trapped particles. The results indicate
that the trapped particle current is negligible. Also note that the trapped
particle currents in the core region and near the edge have opposite
directions, with the edge current being in the co-current direction. This
phenomenon has connection with the so-called ``banana current'', which is an
analogue of the diamagnetic current (the latter is due to the gyro-orbits and
is in the perpendicular direction, whereas the former is due to drift-orbits
and is in the parallel direction{\cite{Peeters_2000}}). The banana current can
be considered part of the bootstrap current{\cite{Peeters_2000}}. In the case
shown in Fig. \ref{23-5-10-p4}, the current direction reverse happens at
$\rho_p \approx 0.6$. The volume integrated trapped particle current is in the
co-current direction, although being very small (only 1\% of the total fast
ion current). Figures \ref{23-5-10-p4}c-d plots the poloidal harmonics of
currents carried by passing ions and trapped ions. For the passing ion
current, the $m = 0$ harmonic is dominant, indicating poloidally uniform of
the current. For the trapped ion current, the $m = 1$ harmonic becomes
dominant in the out region $(\rho_p > 0.6)$, indicating poloidal nonuniform.

\begin{figure}[h]
  \resizebox{0.9\columnwidth}{!}{\includegraphics{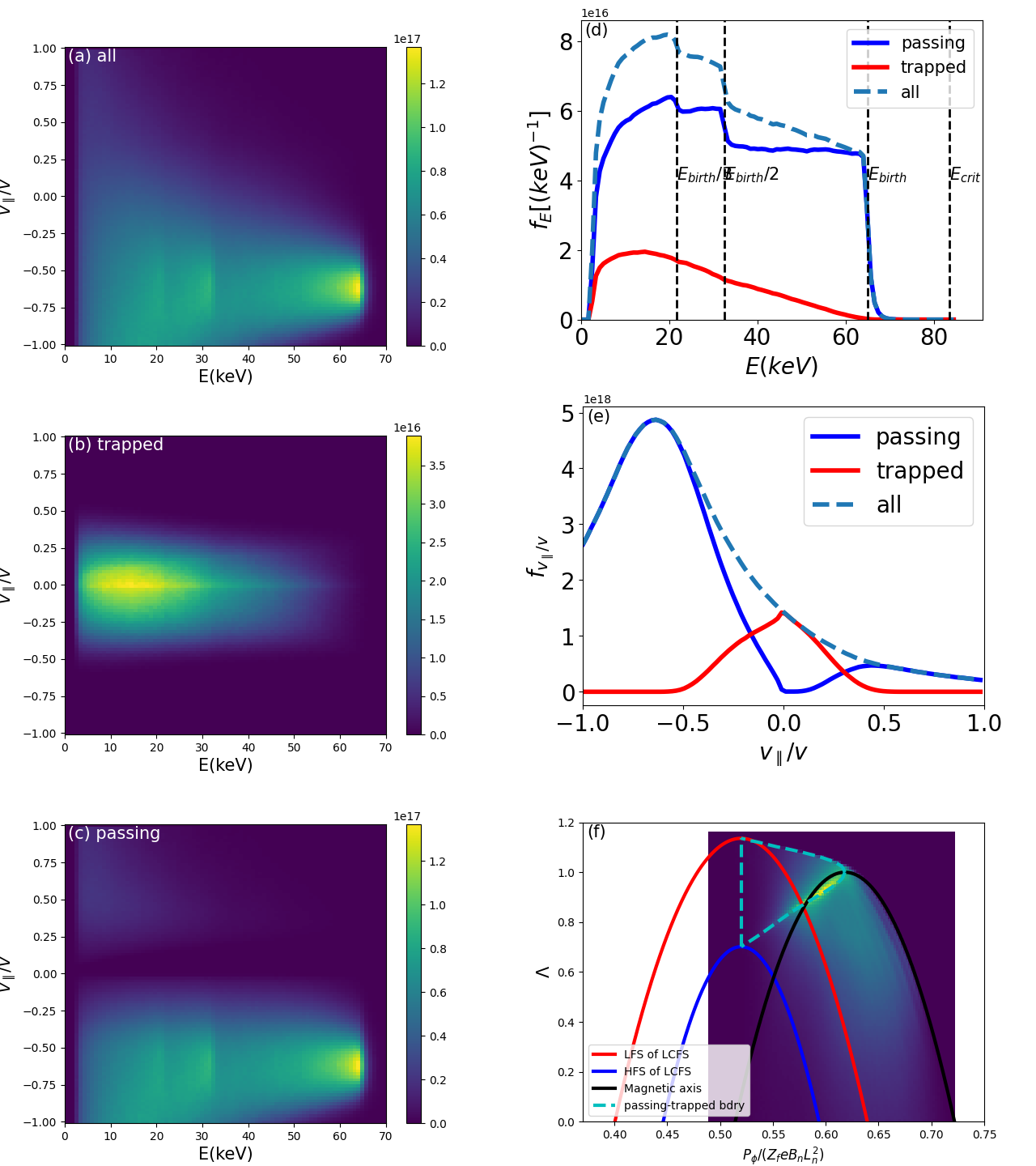}}
  
  \
  \caption{\label{22-10-16-1}Steady-state distribution of NBI fast ions in
  $(E, v_{\parallel} / v)$ (a-c), in $E$ (d), in $v_{\parallel} / v$ (e), and
  in $(P_{\phi}, \Lambda)$ (f). Refer to Fig. \ref{22-10-18-2} for the meaning
  of the various lines in (f). Here $f_E$ is defined by $d N = f_E d E$, where
  $d N$ is the number of particles within the energy interval $d E$. And
  $f_{v_{\parallel} / v}$ and the 2D distributions are defined in a similar
  way. The critical energy for this case $E_{\tmop{crit}} = 83.6 \tmop{keV}$,
  which is larger than the fast ion birth energy.}
\end{figure}

Figures \ref{22-10-16-1}(a)-(e) plot the steady-state fast ion distribution in
$(E, v_{\parallel} / v)$, where $E$ is the kinetic energy and $v_{\parallel}$
is the parallel (to the magnetic field) velocity. Both the 2D distribution and
1D distributions are plotted. We also distinguish between the passing
particles and trapped particles. The results indicate that the trapped
particle distribution is roughly symmetrical in $v_{\parallel} / v$ around
$v_{\parallel} / v = 0$, indicating it carries nearly zero parallel current.
The results also indicate that trapped fast ions are more localized in the low
energy region when compared with passing fast ions which have nearly uniform
energy spectrum. There are three jumps in Fig \ref{22-10-16-1}(d), which
correspond to the NBI source at full, half, and $1 / 3$ of $65 \tmop{keV}$.
Note that collisional energy diffusion makes some particles exceed the full
energy, as is shown in Fig \ref{22-10-16-1}(d).

\

\begin{figure}[h]
  \resizebox{0.9\columnwidth}{!}{\includegraphics{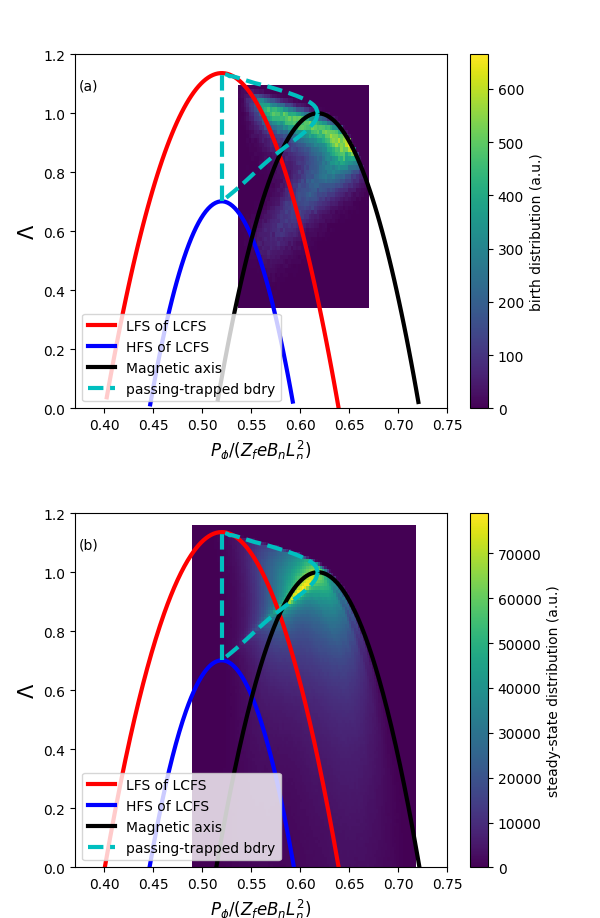}}
  \caption{\label{23-7-19-p3}NBI fast ion birth distribution (a) and
  steady-state distribution (b) in $(P_{\phi}, \Lambda)$ plane for the
  $R_{\tan} = 0.731 m$ beam.}
\end{figure}

Figure \ref{22-10-16-1}(f) plots the steady-state distribution in the
$(P_{\phi}, \Lambda)$ plane. The fast ions seem to reach their peak density
near the passing-trapped boundary. As a comparison, the birth distribution, as
is shown in Fig. \ref{22-10-18-2}, has no obvious structure near the
passing-trapped boundary. The structure of the steady-state distribution near
the passing-trapped boundary is not sensitive to the fast ion birth profile.
For instance, Fig. \ref{23-7-19-p3} considers the \ $R_{\tan} = 0.731 m$ beam
and compares the birth distribution and steady-state one, which shows that the
latter still reach its peak near the passing-trapped boundary.

\subsection{Dependence of fast ion current on plasma current}

\begin{figure}[h]
  \resizebox{0.8\columnwidth}{!}{\includegraphics{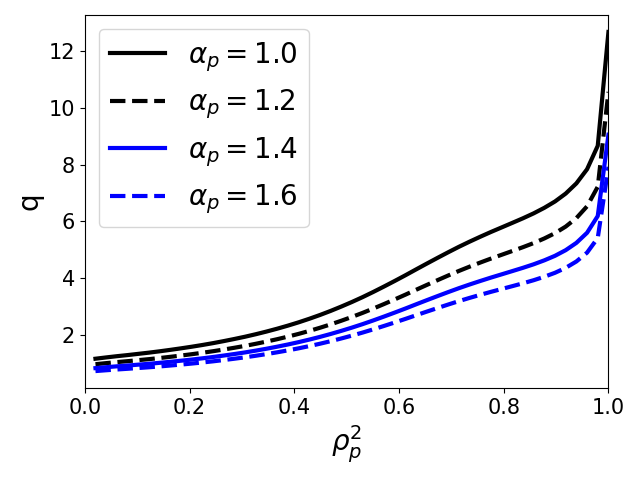}}
  \caption{\label{23-4-13-e1}Safety factor radial profiles for different
  values of $\alpha_p$.}
\end{figure}

Let us examine the dependence of NBCD on the plasma current. We change the
plasma current by multiplying the poloidal magnetic flux $\Psi$ (2D data from
EFIT) by a factor $\alpha_p$. Since the poloidal magnetic field $\mathbf{B}_p$
is related to $\Psi$ via
\begin{equation}
  \mathbf{B}_p = - \frac{1}{R} \frac{\partial \Psi}{\partial Z}
  \hat{\mathbf{R}} + \frac{1}{R} \frac{\partial \Psi}{\partial R}
  \hat{\mathbf{Z}},
\end{equation}
where $\hat{\mathbf{R}}$ and $\hat{\mathbf{Z}}$ are the cylindrical unit
vectors, the above multiplication corresponds to multiplying the
$\mathbf{B}_p$ by $\alpha_p$. Similarly, since the plasma toroidal current
density $J_{\phi}$ is related to $\Psi$ via
\begin{equation}
  J_{\phi} = - \mu_0^{- 1} \frac{1}{R} \frac{\partial^2 \Psi}{\partial Z^2} -
  \mu_0^{- 1} \frac{\partial}{\partial R} \left( \frac{1}{R} \frac{\partial
  \Psi}{\partial R} \right),
\end{equation}
where $\mu_0$ is the vacuum magnetic permeability, the above scaling also
scales the plasma current $I_p$ by $\alpha_p$. We keep the values of the
toroidal magnetic field $B_{\phi}$ fixed when changing $\Psi$. As a result,
the safety factor $q$ is also scaled by a factor of $\alpha_p$. Figure
\ref{23-4-13-e1} plots the safety factor profiles corresponding to different
values of $\alpha_p = I_p / I_{p, \tmop{ref}}$, where $I_{p, \tmop{ref}}$ is
the plasma current in the original configuration.

\begin{figure}[h]
  \resizebox{0.9\columnwidth}{!}{\includegraphics{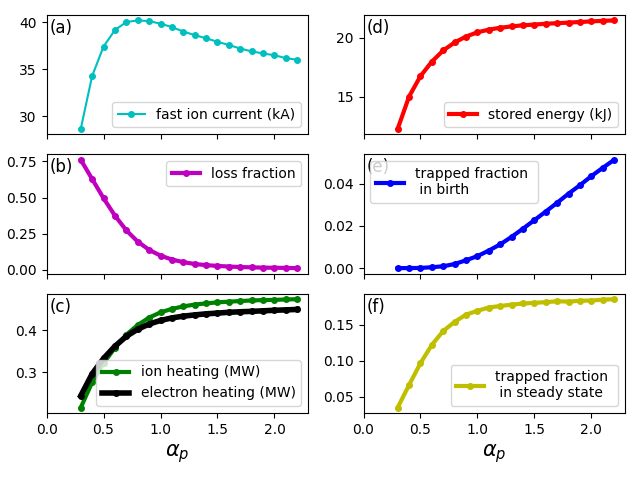}}
  \caption{\label{23-4-23-a1}Fast ion toroidal current (a), edge particle loss
  fraction (b), ion and electron heating power (c), saturated stored energy
  (d), trapped fraction in birth (e), and trapped fraction in steady-state (f)
  as a function of the plasma current $I_p = \alpha_p I_{p, \tmop{ref}}$.}
\end{figure}

Figure \ref{23-4-23-a1}a plots the fast ion current as a function of the
plasma current, which indicates that the dependence is not monotonic: in the
lower $I_p$ regime, the fast ion current $I_f$ increase with $I_p$ increasing
whereas, in the higher $I_p$ regime, $I_f$ decreases with $I_p$ increasing.

The increasing of $I_f$ with $I_p$ increasing in lower $I_p$ regime is
probably due to the improvement of fast ion confinement. The drift orbit width
is inverse proportional to the poloidal magnetic field{\cite{wesson2004}}.
Larger plasma current implies stronger poloidal magnetic field, hence smaller
drift orbit width and thus better confinement of fast ions. The evidence for
this is shown in Fig. \ref{23-4-23-a1}b, which indicates that the fast ion
loss fraction decreases rapidly with the plasma current increasing in the
lower $I_p$ region.

Next, we try to understand why the fast ion current decrease with plasma
current increasing in high $I_p$ regime. The only hint that we can identify is
that the trapped fraction increase with the plasma current. The increase in
trapped fraction happens for both the birth distribution and the steady-state
distribution, as is shown in Fig. \ref{23-4-23-a1}(e) and (f).

\

\

\

\begin{figure}[h]
  \
  
  \resizebox{0.9\columnwidth}{!}{\includegraphics{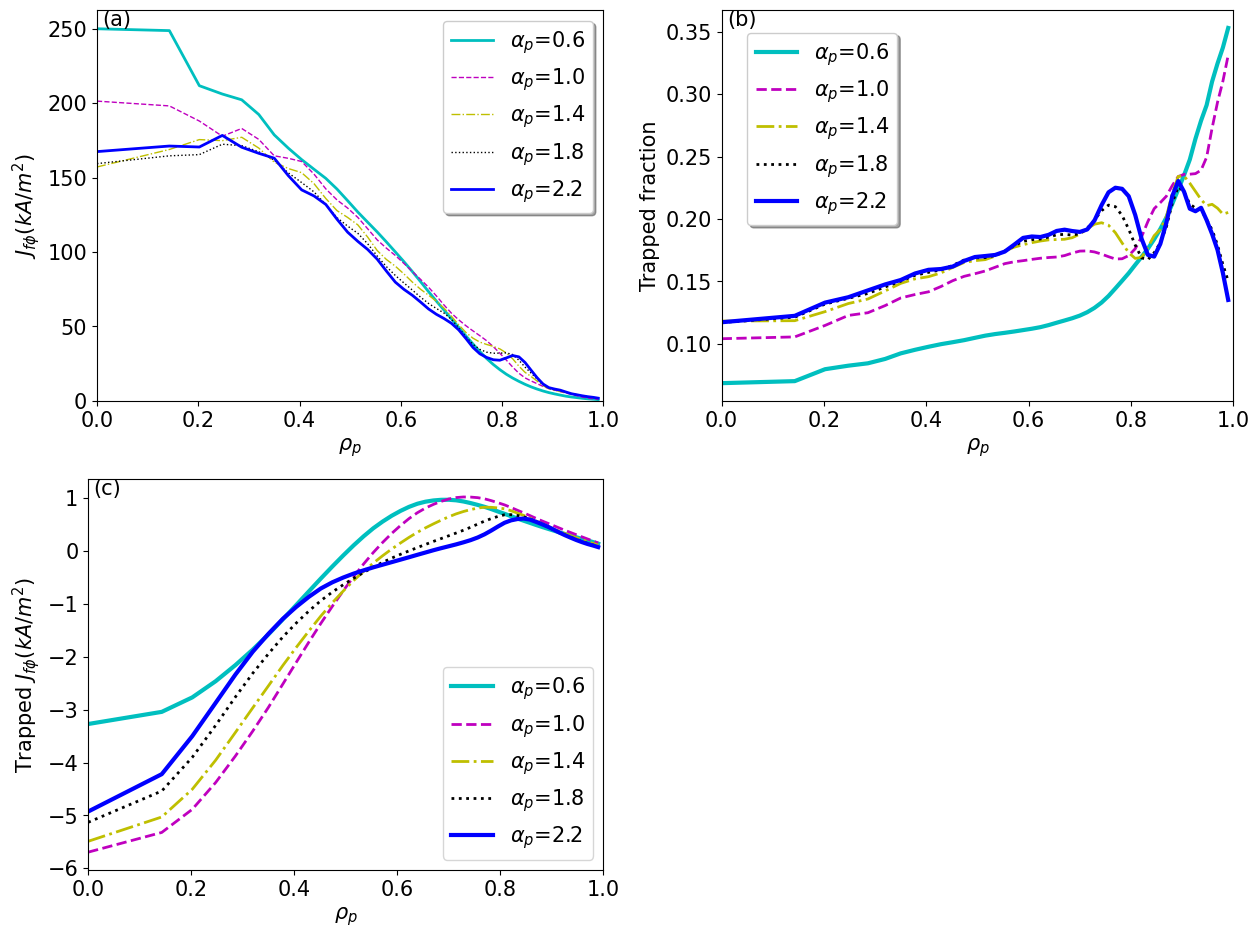}}
  
  \
  \caption{\label{23-5-9-p1}Radial profiles of fast ion toroidal current
  densities (a), trapped fractions (b), and trapped fast ion current densities
  (c) in magnetic configurations with different plasma currents.}
\end{figure}

Figure \ref{23-5-9-p1} plots the radial profiles of fast ion currents, the
trapped fraction, and the current carried by the trapped fast ions, for
various plasma currents. The results indicate again that trapped particles
carry negligible current. Increasing the plasma current makes the trapped
fraction increased across a wide radial region except the edge.

\begin{figure}[h]
  \resizebox{0.95\columnwidth}{!}{\includegraphics{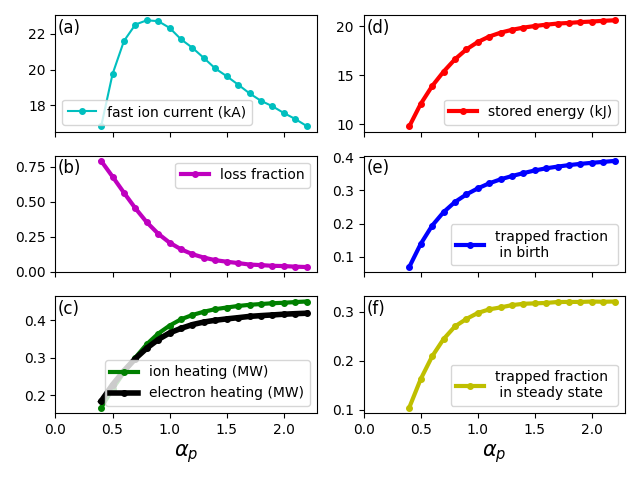}}
  \caption{\label{23-4-24-p5}The same as Fig. \ref{23-4-23-a1} except for a
  smaller tangential radius, $R_{\tan} = 0.731 m$.}
\end{figure}

\begin{figure}[h]
  \resizebox{0.95\columnwidth}{!}{\includegraphics{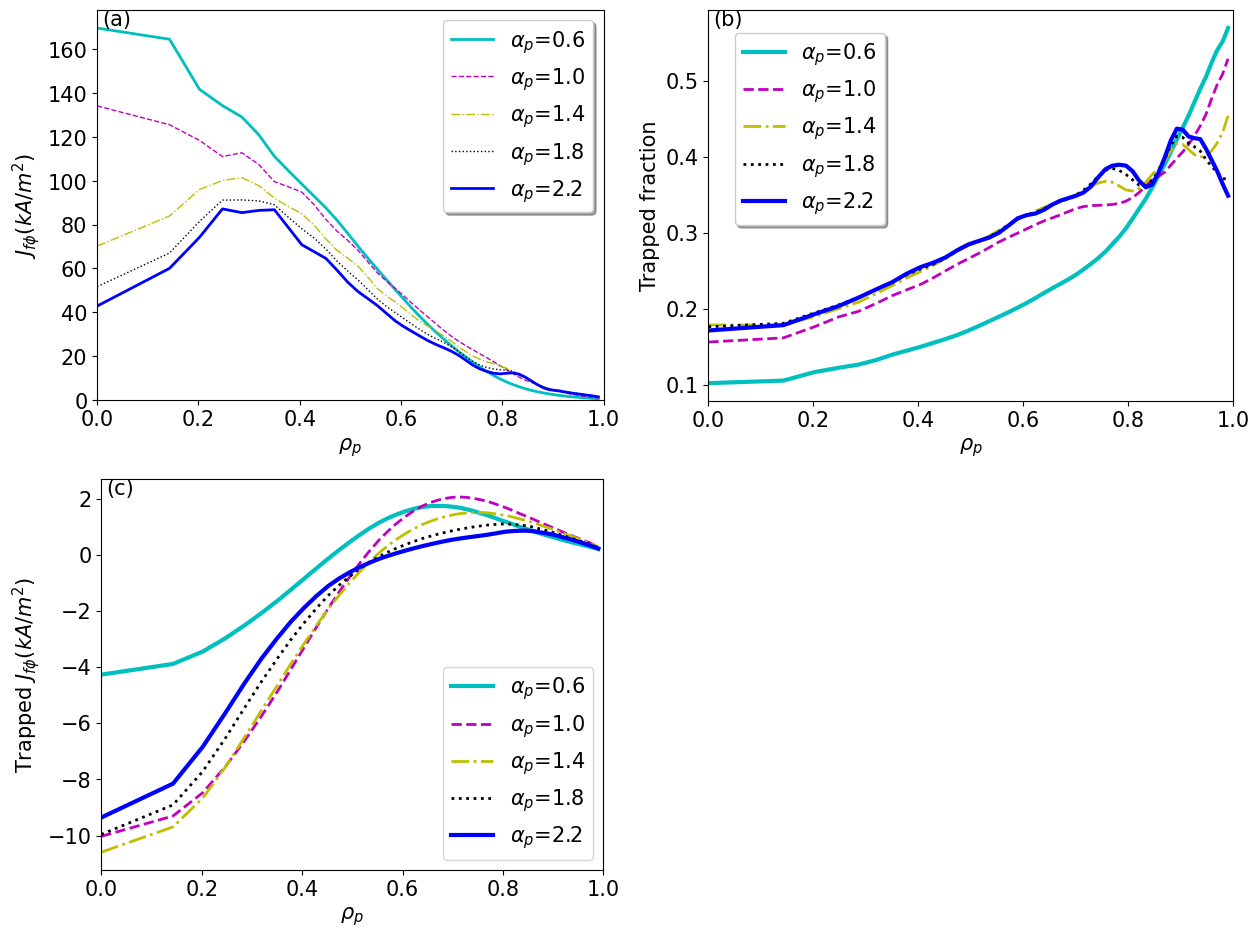}}
  \caption{\label{23-5-10-p1}The same as Fig. \ref{23-5-9-p1} except for a
  smaller tangential radius, $R_{\tan} = 0.731 m$.}
\end{figure}

We also performed similar simulations as above but with a smaller NBI
tangential radius, $R_{\tan} = 0.731 m$. This case has larger trapped particle
fraction. The results are plotted in Figs. \ref{23-4-24-p5} and
\ref{23-5-10-p1}. We observe the same non-monotonic dependence of the fast ion
current on the plasma current. And also the results indicate that the trapped
particles carry nearly zero current.

In the above scanning of plasma current, the density and temperature profiles
are kept fixed. Therefore the force-balance is not guaranteed. This is a
weakness of this work. Similarly, the equilibrium is not re-computed in the
density scanning presented in the next section.

\subsection{Dependence of fast ion current on plasma density}

Next, we examine the plasma density dependence of fast ion current. The
density affects both the ionization process and the fast ion collisional
transport. The ionization process affects the shine-through fraction and ion
birth location. The latter influences the first-orbit loss fraction and the
ratio between trapped and passing fast ions. The collision transport
influences the fast ion distribution in both position space (edge loss) and
velocity space (slowing-down and passing-trapped transition). These effects
may have opposite effects on the fast ion current. Therefore it is not obvious
what is the trend of fast ion current changing with the plasma density. Next,
we examine this trend via simulations.

\begin{figure}[h]
  \resizebox{0.95\columnwidth}{!}{\includegraphics{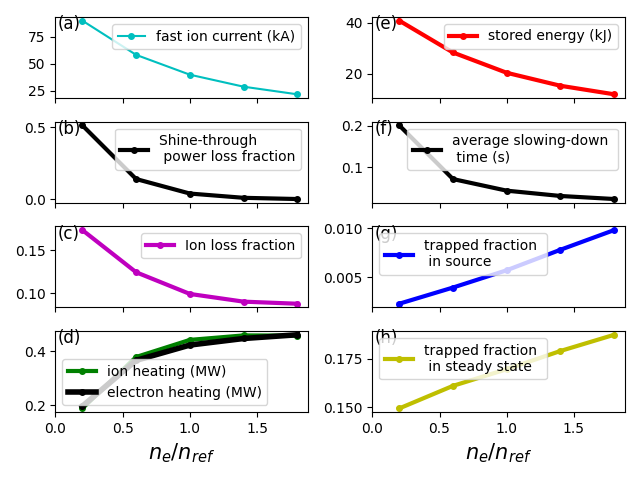}}
  \caption{\label{23-5-10-1}The plasma density dependence of fast ion current
  (a), shine-through loss (b), ion loss fraction (c), ion/electron heating
  power (d), fast ion stored energy (e), slowing-down time (f), and trapped
  fraction in birth distribution (g) and in steady-state distribution (h).}
\end{figure}

We scan the electron density via scaling the profile in Fig. \ref{17-2-22-p1m}
by a factor of values ranging from $0.2$ to 1.8. Fig. \ref{23-5-10-1}a plots
the dependence of the volume integrated fast ion current on the density, which
indicates that the current decreases with increasing of the density. Fig.
\ref{23-5-10-1}b shows that the shine-through loss decrease with the density
increasing (as is expected). Also Fig. \ref{23-5-10-1}c shows that the ion
edge loss fraction decreases with the density increasing. These reduction of
loss is beneficial for current drive since more fast ions can stay in the
plasma to contribute electric current. On the other hand, Fig.
\ref{23-5-10-1}(f-h) shows that, with the density increasing, the slowing-down
time becomes shorter and the trapped particle fraction becomes larger. These
are deleterious effects for current drive. Shorter slowing-down time implies
that fast ions can remain energetic for shorter time and thus less fast ions
can contribute to the fast ion current. Larger trapped fraction means less
particles can efficiently contribute to the current. The results in Fig. \
\ref{23-5-10-1}(a) indicates that the deleterious effects turn out to defeat
those beneficial effects, making the fast ion current decreases with the
density increasing.

\

\begin{figure}[h]
  \resizebox{0.7\columnwidth}{!}{\includegraphics{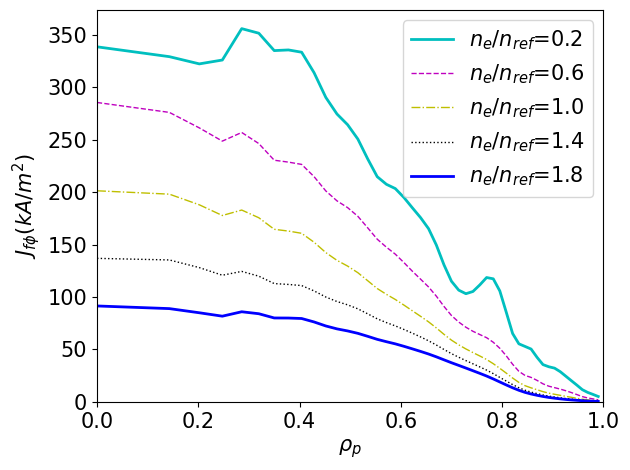}}
  \caption{\label{23-5-10-3}Radial profiles of fast ion current density for
  various plasma density.}
\end{figure}

Figure \ref{23-5-10-3} plots the radial profiles of fast ion current density.
The results indicate that, for all radial positions, the fast ion current
density decreases with the plasma density increasing.

\

\

\subsection{Electron shielding current}

\begin{figure}[h]
  \resizebox{0.7\columnwidth}{!}{\includegraphics{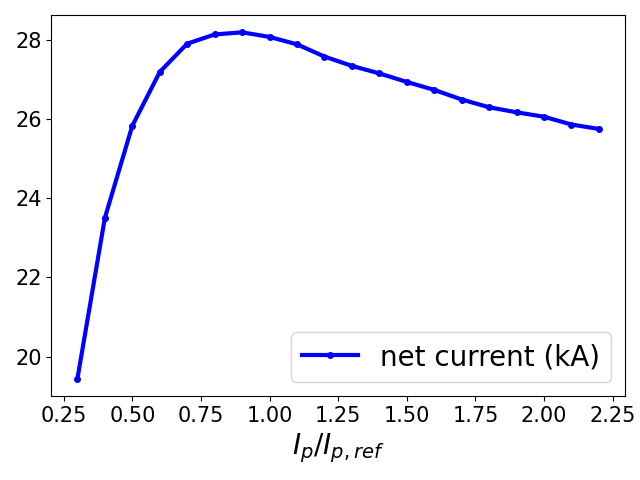}}
  \caption{\label{23-5-12-1}The same as Fig. \ref{23-4-23-a1}a except that
  this is the driven current that includes the electron return current.}
\end{figure}

\begin{figure}[h]
  \resizebox{0.7\columnwidth}{!}{\includegraphics{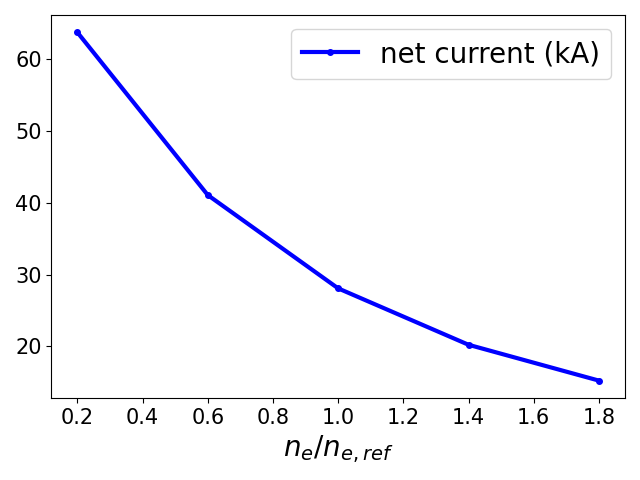}}
  
  \
  \caption{\label{23-5-12-2}The same as Fig. \ \ref{23-5-10-1}a except that
  this is the net current rather than the fast ion current.}
\end{figure}

Taking into account the electron shielding effect on the fast ion current does
not qualitatively change the dependence of the driven current on the plasma
current, i.e., net current follows a similar trend as the fast ion current, as
is shown in Fig. \ref{23-5-12-1} and \ref{23-5-12-2}.

\

\

\begin{figure}[h]
  \resizebox{0.95\columnwidth}{!}{\includegraphics{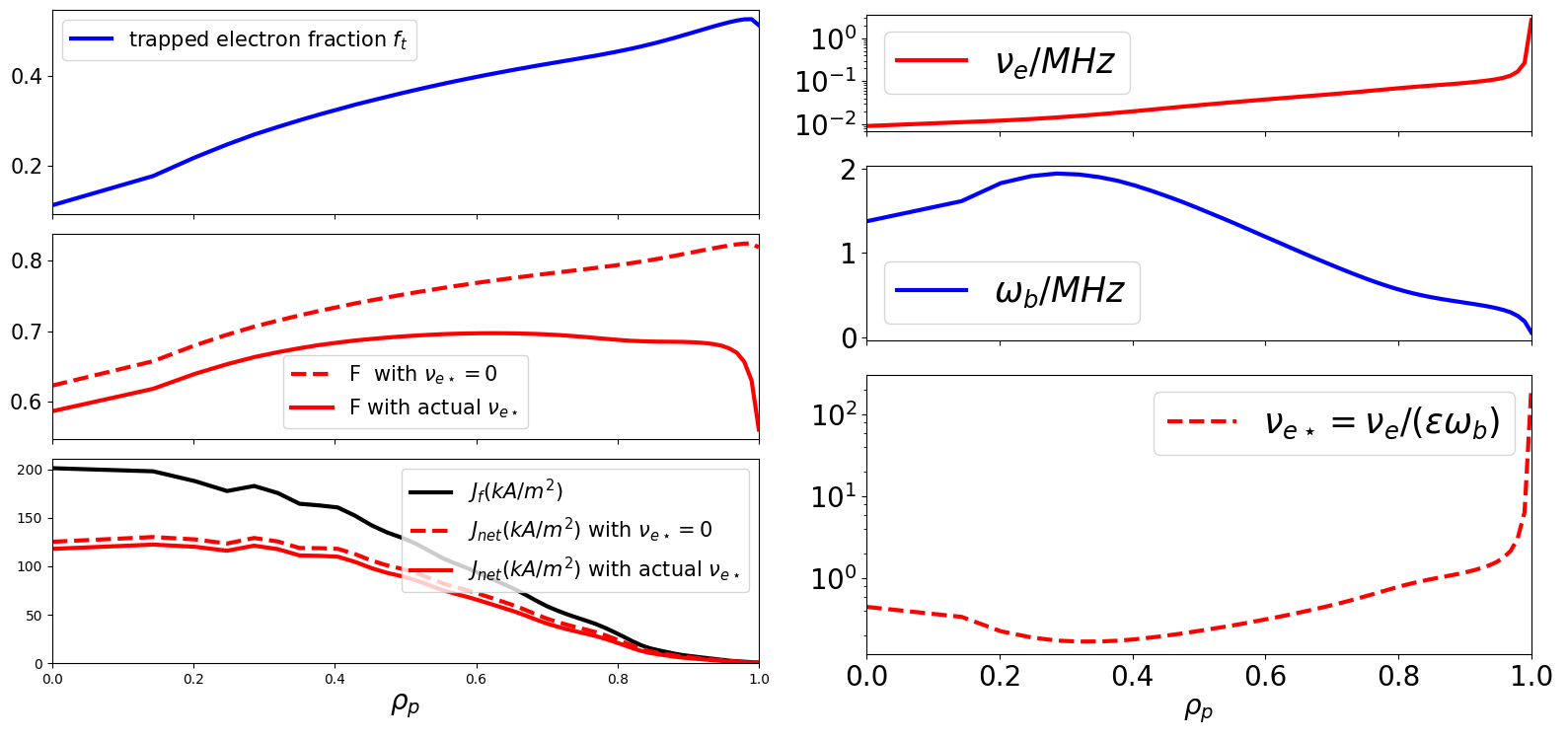}}
  \caption{\label{22-10-26-3}Radial profiles of effective trapped electron
  fraction $f_t$, the shielding factor $F$, the pure fast ion current density
  $J_f$, the net current density $J_{\tmop{net}} = J_f F$, the electron
  collision frequency $\nu_e$, the thermal electron bounce frequency
  $\omega_b$, and the normalized electron collision frequency $\nu_{e \star}$.
  Some quantities are undefined at the magnetic axis. Their values are
  obtained by interpolation.}
\end{figure}

Figure \ref{22-10-26-3} plots the radial profiles of various quantities
related to the driven current, namely, the pure fast ion current density
$J_f$, the electron shielding factor $F$, the net current density
$J_{\tmop{net}} = J_f F$, the effective trapped electron fraction $f_t$, \ the
electron collision frequency $\nu_e$, the thermal electron bounce frequency
$\omega_b$, and the normalized electron collision frequency $\nu_{e \star}$.
(The details of these quantities are given in Appendix \ref{22-10-26-p6}.) The
formulas for the shielding effect used here are valid for general tokamak
equilibria and arbitrary collisionality
regime{\cite{Honda_2012,youjunhu2012}}, where the equilibrium shaping effects
are included via the effective trapped fraction $f_t$.

\

As a comparison, we also plot the shielding factor $F$ for the case of
$\nu_{e \star} = 0$ and the resulting net current. The results show that the
$\nu_{e \star} = 0$ approximation slightly overestimates the net current.

\section{Summary}

Simulations of neutral beam current drive on the EAST tokamak were performed,
providing detailed information about the fast ion distribution in both real
space and velocity space. We distinguish between current carried by passing
particles and that carried by trapped particles, and found that trapped ions
carry co-current current near the edge and counter-current current near the
core. However, the magnitude of the trapped ion current is one order smaller
than that of the passing ions, making their contribution to the fast ion
current negligible.

We examine the dependence of the fast ion current on the plasma current $I_p$.
With $I_p$ increasing, the drift orbit width decreases, which helps reduce the
first-orbit loss and collisional transport, and thus is beneficial for fast
ion confinement. This mechanism makes the fast ion current increase with $I_p$
in the low $I_p$ regime. But for high $I_p$ regime, the effect of trapped
fraction increasing becomes dominant. Since the trapped fast ions carry nearly
zero current, the increasing of trapped fraction then implies the decreasing
of fast ion current.

The fast ion current decreases with the increase of plasma density $n_e$.
This dependence is mainly determined by the variation of the slowing-down time
with $n_e$, which is already well known and is confirmed in our specific
situation.

\section{ACKNOWLEDGMENTS}

Youjun Hu thanks \ Youwen Sun, G. S. Xu, Yingfeng Xu, Lei Ye, Yifeng Zheng,
and Baolong Hao for useful discussions, and thanks Xiaotao Xiao for careful
reading of the first version of this paper. The manuscript of this paper was
written in GNU TeXmacs, a free structured WYSIWYG editor for
scientists{\cite{texmacs}}. Numerical computations were performed on Tianhe at
National SuperComputer Center in Tianjin, Sugon computing center in Hefei, and
the ShenMa computing cluster in Institute of Plasma Physics, Chinese Academy
of Sciences. This work was supported by Users with Excellence Program of Hefei
Science Center CAS under Grant No. 2021HSC-UE017, by Comprehensive Research
Facility for Fusion Technology Program of China under Contract No.
2018-000052-73-01-001228, and by the National Natural Science Foundation of
China under Grant No. 11575251.

\subsection{Conflict of interest}

The authors have no conflicts to disclose.

\section{DATA AVAILABILITY}

The data that support the findings of this study are available at
https://github.com/Youjunhu/TGCO, and are licensed under the GNU General
Public License v3.0.

\appendix\section{Continuous beam injection}\label{22-11-1}

\begin{figure}[h]
  \resizebox{0.8\columnwidth}{!}{\includegraphics{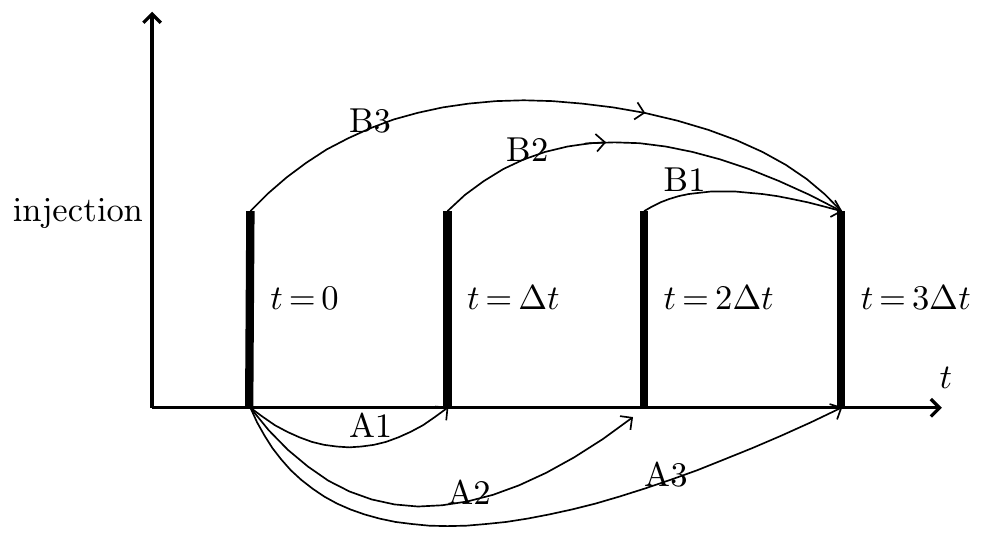}}
  
  \
  \caption{\label{2020-1-6-a1}An efficient way of simulating multiple
  injections. The contribution of injections at $t = j \Delta t$ with $j = 0,
  1, 2$ to the fast ion distribution at $t = 3 \Delta t$ can be obtained by
  following the time history of the particles injected at $t = 0$ and
  recording their contribution to the fast ion distribution at $t = \Delta t,
  2 \Delta t, 3 \Delta t$, indicated respectively by $A 1$, $A 2$, and $A 3$.
  Then it is ready to see that $B 1 = A 1$, $B 2 = A 2$, and $B 3 = A 3$.
  Therefore the contributions of the multiply injections can be inferred from
  the time history of a single injection.}
\end{figure}

To obtain the steady state of fast ions, we need include the continuous beam
source. A straightforward Monte-Carlo implementation of this continuous
injection would be to introduce new Monte-Carlo markers to represent the newly
injected physical particles at each time step. This method is computationally
expensive. For a time-independent background plasma, there is an efficient
method that involves only a single injection and then utilizes the time shift
invariant to infer the contribution of all the other injections. This method
is illustrated in Fig. \ref{2020-1-6-a1}.

\

The above method works only for a time-independent background plasma and
constant beam power. For time-dependent background plasma, re-injecting new
Monte-Carlo markers seems to be the only method available. The present work
considers a time-independent background plasma with a constant beam power and
use the above efficient method.

\section{Collision model}\label{22-11-2-a1}

In the zero drift-orbit width approximation, the time it takes for a fast ion
of velocity $v_1$ to be slowed down to $v_2$ by the collision friction with
the background ions and electrons is given by{\cite{wesson2004}}
\begin{equation}
  \label{22-9-17-p1} \tau_s = \frac{\tau_{s e}}{3} \ln \left[ \frac{1 + \left(
  \frac{v_c}{v_1} \right)^3}{\left( \frac{v_2}{v_1} \right)^3 + \left(
  \frac{v_c}{v_1} \right)^3} \right],
\end{equation}
where $v_c$ is the critical velocity defined by
\begin{equation}
  \label{19-12-24-1} v_c = \left( \frac{3 \sqrt{\pi}}{4}  \frac{m_e}{n_e}
  \sum_i  \frac{n_i Z_i^2}{m_i} \right)^{1 / 3} \sqrt{\frac{2 T_e}{m_e}},
\end{equation}
$\tau_{s e} = 1 / \nu_{s e}$ with $\nu_{s e}$ defined by
\begin{equation}
  \label{20-1-17-a1} \nu_{s e} = \frac{4}{3 \sqrt{\pi}} \frac{m_f}{m_e} 
  \frac{\Gamma^{f / e}}{(2 T_e / m_e)^{3 / 2}},
\end{equation}
which is the slowing-down rate due to background electrons,
\begin{equation}
  \Gamma^{f / e} = \frac{n_e Z_f^2 e^4}{4 \pi \epsilon_0^2 m_f^2} \ln \Lambda,
\end{equation}
$\ln \Lambda$ is the Coulomb logarithm ($\ln \Lambda = 24 - \ln \left( 10^{-
6} \sqrt{n_e} / T_e \right) \approx 16.97$ is used in this work, where $n_e$
and $T_e$ are in units of $m^{- 3}$ and $\tmop{keV}$, respectively).

\begin{figure}[h]
  \resizebox{0.8\columnwidth}{!}{\includegraphics{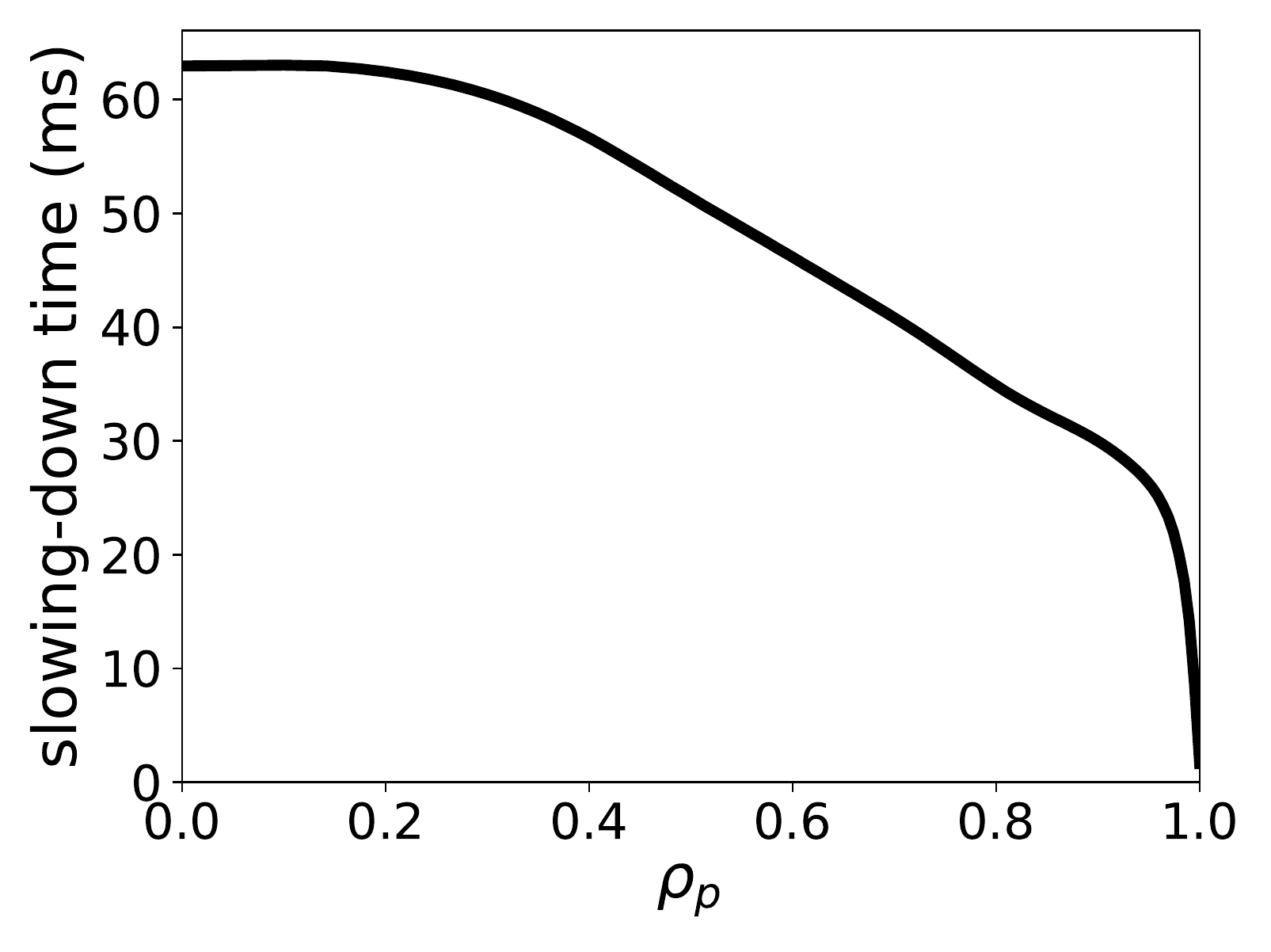}}
  \caption{\label{22-10-26-p1}Radial profiles of slowing-down time for fast
  ions of kinetic energy $65 \tmop{keV}$ (assuming zero drift-orbit width) to
  slow down to the cutoff energy $2 T_i (0)$ in the equilibrium specified in
  Fig. \ref{17-2-22-p1m}.}
\end{figure}

Both $\tau_{s e}$ and $v_c$ have a radial dependence via their dependence on
the plasma density and temperature. The radial profile of $\tau_s$ for the
plasma specified in Fig. \ref{17-2-22-p1m} is plotted in Fig.
\ref{22-10-26-p1} for fast ions of $65 \tmop{keV}$ to be slowed down to a
cutoff velocity (chosen as $m v_2^2 / 2 = 2 T_i (0)$ in this article). The
result shows that typical slowing-down time in the core region is about $60
\tmop{ms}$.

\

The formula (\ref{22-9-17-p1}) only includes the slowing-down, neglecting
pitch-angle scattering and energy diffusion. For realistic simulations that
include pitch-angle scattering, energy diffusion, finite-orbit-width effect,
and non-uniform plasma profiles, we need to use numerical integration to
determined the slowing-down process of fast ions. The Monte-Carlo algorithm
used in this work for collision of fast ions with background electrons and
ions is specified in Ref. {\cite{todo2014}}, where the pitch-angle variable
$\lambda = v_{\parallel} / v$ and velocity $v$ are altered at the end of each
time step according to the following scheme:
\begin{equation}
  \lambda_{\tmop{new}} = \lambda (1 - \nu_d \Delta t) \pm \sqrt{(1 -
  \lambda^2) \nu_d \Delta t},
\end{equation}
and
\begin{eqnarray}
  v_{\tmop{new}} & = & v - \nu_s \Delta t v \left( 1 + \frac{v_c^3}{v^3}
  \right) \nonumber\\
  & + & \frac{\nu_s \Delta t}{m_f v} \left[ T_e - \frac{1}{2} T_i \left(
  \frac{v_c}{v} \right)^3 \right] \nonumber\\
  & \pm & \sqrt{\frac{\nu_s \Delta t}{m_f} \left[ T_e + T_i \left(
  \frac{v_c}{v} \right)^3 \right]}  \label{21-3-29-a1}
\end{eqnarray}
where the second and third line correspond to the energy diffusion, $\pm$
denotes a randomly chosen sign with equal probability for plus and minus,
$\Delta t$ is the time step, $\nu_d$ is the pitch-angle scattering rate given
by
\begin{equation}
  \label{16-9-1-p1} \nu_d = \left( \frac{Z_{\tmop{eff}}}{v^3} \right)
  \frac{n_e Z_f^2 e^4 \ln \Lambda}{4 \pi \varepsilon_0^2 m_f^2},
\end{equation}
where $Z_f$ is the charge number of fast ions, $\nu_s$ is the slowing down
rate given by Eq. (\ref{20-1-17-a1}).

\section{Formulas for electron shielding effect}\label{22-10-26-p6}

The electron collision time (characterizing electron collision with ions)
is{\cite{wesson2004}}
\begin{equation}
  \tau_e = \frac{12 \pi^{3 / 2}}{\sqrt{2}} \dfrac{\varepsilon_0^2  \sqrt{m_e}
  T_e^{3 / 2}}{n_i Z_i^2 e^4 \ln \Lambda_e},
\end{equation}
where $\varepsilon_0$ is the vacuum permittivity, $\ln \Lambda_e$ is the
Coulomb logarithm for electron-ion collision ($\ln \Lambda_e = 15.2 - 0.5 \ln
(n_e / 10^{20}) + \ln (T_e)$ is used in this work, where $n_e$ and $T_i$ are
in units of $m^{- 3}$ and $\tmop{keV}$, respectively). The dimensionless
electron collision parameter $\nu_{e \star}$ is defined by
\begin{equation}
  \label{22-10-14-1} \nu_{e \star} = \frac{\nu_e}{\varepsilon \omega_{b e}},
\end{equation}
where $\nu_e = 1 / \tau_e$, $\varepsilon = r / R_0$ is the local inverse
aspect ratio, $r = (R_{\max} - R_{\min}) / 2$, $R_0 = (R_{\min} + R_{\max}) /
2$, $R_{\min}$ and $R_{\max}$ are, respectively, the maximal and minimal
values of the cylindrical coordinate $R$ on a magnetic surface, $\omega_{b e}$
is the typical bounce (angular) frequency of thermal electrons, which is given
by (in the approximation of zero-orbit-width and deeply trapped electrons)
\begin{equation}
  \omega_{b e} = \frac{\sqrt{2 T_e / m_e}}{q R_0}  \left(
  \frac{\varepsilon}{2} \right)^{1 / 2},
\end{equation}
where $q$ is the safety factor of the magnetic surface. Using this, Eq.
(\ref{22-10-14-1}) is written as
\begin{equation}
  \label{4-12-1} \nu_{e \star} \equiv \frac{\nu_e}{\varepsilon^{3 / 2}
  \sqrt{T_e / m_e} / (q R_0)} .
\end{equation}
Due to the electron shielding effect, the ratio of the net current to the fast
ion current is given by
\begin{equation}
  \label{11-8-p1} F \equiv \frac{\langle j_{\parallel} B \rangle}{\langle j_{f
  \parallel} B \rangle} = 1 - \frac{Z_f}{Z_{\tmop{eff}}} (1 - \mathcal{L}_{31}
  ),
\end{equation}
where $\langle \ldots \rangle$ is the magnetic surface averaging, \
$\mathcal{L}_{31}$ is the bootstrap current coefficient before the electron
density gradient. The formula of $\mathcal{L}_{31}$ given by Sauter {\tmem{et
al.}}{\cite{sauter1999}} is
\begin{eqnarray}
  \mathcal{L}_{31} & = & \left( 1 + \frac{1.4}{Z_{\tmop{eff}} + 1} \right) X -
  \frac{1.9}{Z_{\tmop{eff}} + 1} X^2 \nonumber\\
  & + & \frac{0.3}{Z_{\tmop{eff}} + 1} X^3 + \frac{0.2}{Z_{\tmop{eff}} + 1}
  X^4,  \label{11-8-1}
\end{eqnarray}
with
\begin{equation}
  X = \frac{f_t}{1 + (1 - 0.1 f_t) \sqrt{\nu_{e \star}} + 0.5 (1 - f_t) \nu_{e
  \star} / Z_{\tmop{eff}}},
\end{equation}
$f_t$ is the effective trapped fraction{\cite{Hirshman_1981,linliu1995}},
\begin{equation}
  \label{11-8-2} f_t = 1 - \frac{3}{4} \left\langle \frac{B^2}{B_{\max}^2}
  \right\rangle \int_0^1 \frac{\lambda d \lambda}{\left\langle \sqrt{1 -
  \lambda B / B_{\max}} \right\rangle},
\end{equation}
where $B_{\max}$ is the maximal value of magnetic field strength on a magnetic
surface.

\end{document}